\documentclass[aps,elsart,onecolumn,groupedaddress]{revtex4}
\usepackage{longtable}
\usepackage{graphicx}% Include figure files
\usepackage{dcolumn}% Align table columns on decimal point
\usepackage{bm}% bold math
\begin{document}
%Title of paper
\title{A spatial interpretation of emerging superconductivity in lightly doped cuprates}

\author{Guy Deutscher} \email{guyde@post.tau.ac.il}
\affiliation{School of Physics and Astronomy, Tel-Aviv University,
Tel Aviv, 69978, Israel}
\author{Pierre-Gilles de Gennes}
\affiliation{Institut Curie, Recherche, 26, rue d'Ulm, 75005 Paris,
France}


\begin{abstract}
The formation of domains comprising alternating 'hole rich' and
'hole poor' ladders recently observed by Scanning Tunneling
Microscopy by Kohsaka et al., on lightly hole doped cuprates, is
interpreted in terms of an attractive mechanism which favors the
presence of doped holes on Cu sites located each on one side of an
oxygen atom. This mechanism leads to a geometrical pattern of
alternating hole-rich and hole-poor ladders with a periodicity equal
to 4 times the lattice spacing in the CuO plane, as observed
experimentally. To cite this article: G. Deutscher, P.-G. de Gennes,
C. R. Physique 8 (2007).
\end{abstract}

% insert suggested PACS numbers in braces on next line
% insert suggested keywords - APS authors don't need to do this
%\keywords{}

%\maketitle must follow title, authors, abstract, \pacs, and \keywords
\maketitle
\section {Introduction}
Recent imaging of Scanning Tunneling Spectroscopy (STM) of lightly
doped superconducting cuprates has revealed the existence of
rectangular domains of width 4a, where a is the lattice spacing in
the CuO plane \cite{ref1}. Inside each domain the carrier
concentration is non-uniform, with a sharp contrast between a
central ladder consisting of a column of oxygen atoms and the two
neighboring Cu-–O–-Cu columns where the carrier concentration is
high, while it is low on similar ladders at the edges of the domain.
The general aspect of the images shows domains having two possible
orientations at right angles to each other, with less well organized
domains spread randomly across the surface. Tunneling
characteristics measured at sites belonging to the central ladder
show at low bias a structure reminiscent of a superconducting
density of states, with pronounced peaks particularly on the
Cu-–O–-Cu columns, while at the edge ladders only a small
conductance dip is visible.

In this communication we wish to point out that this geometrical
pattern can be easily understood if one assumes the existence of a
mechanism that favors the presence of a pair of doped holes on
Cu–-O–-Cu, or perhaps O–-Cu-–O–-Cu–-O segments, accompanied by a
small contraction of the Cu–-O distances. This mechanism can be due
to an increase of the transfer integral t$_{OCu}$. t$_{OCu}$ depends
critically on the overlap between Cu(d) orbitals and O(p) orbitals.
If the energy gained by contraction is sufficient, a bound state of
the hole pair can be formed.

\section {The model}
Our interpretation of the pattern observed in \cite{ref1} proceeds
in three steps: (i) Pair formation; (ii) Formation of holerich and
hole-poor regions; (iii) Pair propagation.

\begin{figure}
  \includegraphics[width=0.85\hsize]{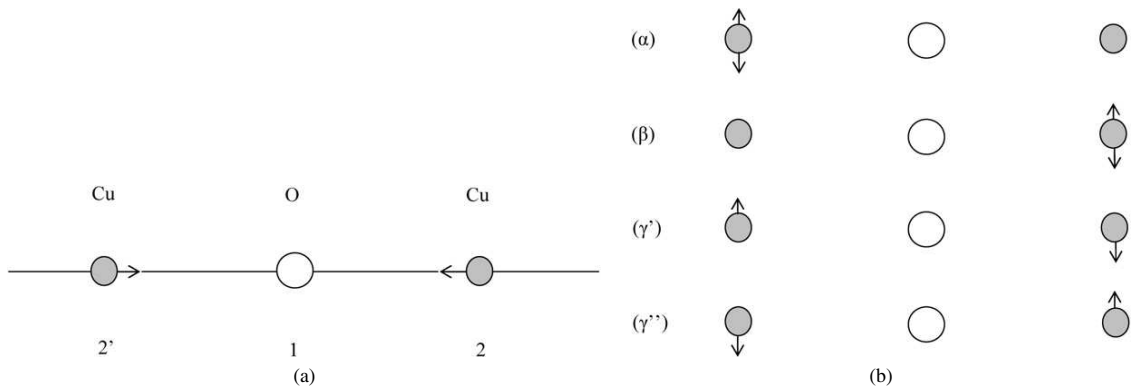}\\
  \caption{(a) A two hole pair on two copper atoms (2) and (2'),
around one oxygen atom (1). We assume here that the two coppers move
in when each of them carries a hole: the transfer integral between 2
and 2' is increased. (b) The four states participating in the
construction of a spin singlet pair.}
\end{figure}

\begin{figure}
\includegraphics[width=0.7\hsize]{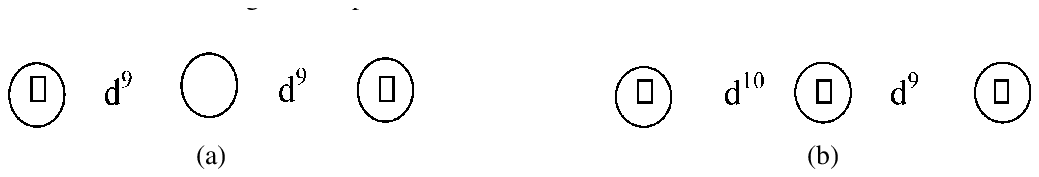} \caption{(a) An oxygen centered O–-Cu-–O-–Cu-–O segment with holes
located on the edge oxygens. Energy is lowered by admixture with
states as shown in Fig. 2b.}
\end{figure}
\subsection {Pair formation}
We discuss on Fig. 1 the binding energy of a hole pair on a
Cu-–O–-Cu segment. We estimate it by a variational function
containing four states $\alpha, \beta, \gamma'$  and $\gamma''$ as
defined in Fig. 1(b). We are interested here in the spin singlet,
for which the amplitudes are even ($\alpha=\beta,
\gamma'=\gamma''$). The unperturbed energy of ($\alpha$) and
($\beta$) is \emph{U} and is large. The eigenvalue equation for the
energy $\epsilon$ is then:
\begin{equation}
\epsilon(\epsilon-U)=4(t_{OCu})^2
\end{equation}
This leads (for large $U$) to an energy:
\begin{equation}
\epsilon_{pair}=-4(t_{OCu})^2/U
\end{equation}
(We are indebted to W. Harrison for pointing out the factor 4
showing up in the singlet energy.) We must compare this to the
energy of two independent holes, both at the bottom of the band,
$2\epsilon_0$, where $\epsilon_0=4t_0$ is the half band width. The
pairs win if:
\begin{equation}
t_{OCu}=(Ut_0/2)^{1/2}
\end{equation}
On the whole, this idea is very tentative but it has one merit: the
ratio $t_{Ocu}/t_0$ is very sensitive to the (Cu$d$)(O $p$) overlap;
this could explain why ions which are isoelectronic to Cu cannot
compete with the cuprates.

There is, however, a difficulty with holes residing on copper atoms,
in view of the high ionization potential of the Cu$^{++}$ ion
\cite{ref2}. Holes may reside preferentially on oxygen atoms. This
can be achieved if we consider a O–-Cu-–O-–Cu-–O segment, with holes
on the edge oxygen atoms (Fig. 2). Here, energy can be gained by
admixture with states where a hole is transferred from one of the
coppers to the central oxygen. The energy gained is still
proportional to $(t_{OCu})^2$.
\subsection {Formation of hole-rich and hole-poor regions}
So let us assume that two holes have formed a bound state, and let
us examine the consequences for the neighboring sites, as shown in
Fig. 3. Following Kohsaka et al. we label 1 the site of the central
oxygen atom; 2 and 2' the neighboring copper sites; 3 and 3' the
following copper sites, followed by the oxygen sites 4 and 4'. Since
the contraction, or dimerisation, of the segment (2–-2') goes hand
in hand with a local increase of the hole density, it follows that
the hole density on neighboring sites such as copper sites 3 and 3'
is decreased. A charge density wave can be triggered by local pair
formation, dimerisation and the charge density wave being coupled
phenomena, the two reinforcing each other. We then arrive at the
following sequence: the central Cu-–O-–Cu bond (2'-–1-–2) is
hole-rich, the neighboring bonds (3-–4) and (4'–-3') are hole-poor.
Over a distance of 4a, there are two hole-rich and two hole-poor Cu
sites. Defects, due primarily to independent nucleation of other
dimers at right angles with the one under consideration, will pin
the charge density wave. The pair on 2'-–2 cannot propagate along
its axis. These conclusions still hold if the hole pair is localized
on a five site segment.
\subsection {Pair propagation}
There is nothing to prevent pair propagation in the perpendicular
direction, which will reduce the kinetic energy. This will lead to
the formation of domains consisting of one central hole-rich and two
lateral hole-poor ladders, the central ladder being centered on an
oxygen column. Pair propagation on the central ladder can lead to
incipient superconductivity. A periodicity of 4a (1), as has been
observed, with alternating conducting (and may be superconducting),
and insulating (and may be antiferromagnetic) domains can be
favorable in some range of doping, see below a discussion of this
point.

\begin{figure}
  \includegraphics[width=0.5\hsize]{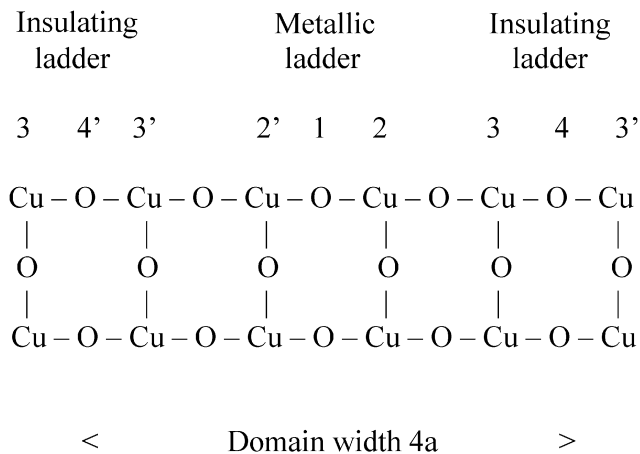}\\
  \caption{The pattern of hole concentration in a domain consisting
  of one high concentration central ladder centered on a column of
  oxygen atoms with two low concentrations ladders on each side. The
  pattern leads to a periodicity of 4 times the lattice parameter of the
  CuO plane.}
\end{figure}

Domain formation is the result of a nucleation process of the first
bound state. This nucleation process occurs randomly along the (10)
and (01) directions, resulting in domains at right angles randomly
distributed, again as observed. There is no long range order in this
pattern.
\section{Discussion}
The essential difference between the proposed model and that of
stripes \cite{ref3,ref4,ref5} is that ours is not a purely
electronic one reflecting the competition between AF and
superconducting orders, where the lattice plays no role (or only a
very secondary one). Rather, the contraction of Cu–-O bonds plays
here a central role. Both models predict oxygen centered structures,
in agreement with experiments \cite{ref1,ref6}, but differ in some
important ways. While stripes are basically one dimensional (namely
their length should be much longer than their width), our model
being based on a local bond contraction leads to a pattern of
domains, at right angles to each other, that are not necessarily
very long, in better agreement with what is observed \cite{ref1},
and with several pieces of experimental evidence for an Electronic
Cluster Glass in underdoped samples (\cite{ref7} and references
therein), rather than one dimensional stripes. Lattice disorder
below the temperature of formation of the domains is a natural
consequence of our local pair formation model, since bond
contraction is local and occurs randomly in two different
directions. Lattice disorder has indeed been found to increase below
a temperature of about 150 K, which we surmise is that of formation
of the domains \cite{ref8,ref9}.

The hole doping level of the samples studied by Kohsaka et al. is of
the order of 0.1 per Cu site. The minimum number of Cu sites
necessary to have at least one hole pair per domain is then 20,
which means that slightly elongated domains as observed do contain a
few hole pairs. Conduction can take place along the central ladder.
If we assume that the hole concentration is peaked on the central
ladder and is negligible on the edge ladders, the hole concentration
on the central ladder is then of the same order as the average
concentration at optimum doping. The edge ladders are then at a
concentration near that of the pure antiferromagnetic phase. Barring
complications due to the one-dimensional character of the central
ladders, one may also expect the formation of a superconducting gap
of the same order of magnitude as that of the gap at optimum doping,
i.e. of a few 10 meV, again as observed.

The emerging picture of a lightly doped cuprate is then that it is
composed of conducting and superconducting ladders weakly coupled
together, either laterally through insulating ladders or at right
angles. Conduction and superconduction on the macroscopic scale will
necessarily require coupling between ladders oriented at right
angles. Since translational and rotational symmetry are broken in
the ladder pattern, which is highly disordered, it is not obvious
that quasi-particles can exist with a well defined wave vector
formed in lightly doped cuprates. Indeed, the normal state is known
to be weakly insulating. If some degree of coherence can be
achieved, one may expect that it will be for quasi-particles having
their momentum at 45 degrees from the ladders directions, which is
the (11) orientation and equivalent. Clearly, there is no
possibility of propagation along the (10) and equivalent directions
because of the interruption of the domains. This is in agreement
with the known fact that lightly doped cuprates have an incomplete
Fermi surface consisting of small arcs around the (11) directions
\cite{ref10}.We note that strong renormalization effects near the
Fermi level around the (11) direction over an energy range of the
order of 100 meV have recently been attributed to electron–phonon
interaction \cite{ref11}.

Superconductivity on themacroscopic scale can be very different from
that in the individual domains, again because it requires coupling
between domains that are at right angles from each other. A gap
anisotropy is to be expected, with the strongest gap values being
along the (10) and equivalent directions. It may be that this
coupling is also at the origin of the d-wave symmetry of the
macroscopic superconducting order parameter.
\par
\textbf{Note added by Guy Deutscher at the time of submission of the
final version} Pierre-Gilles de Gennes passed away a few days before
we were to have a final discussion of the manuscript. He had left
clear instructions that I was to take care of the final version. It
is under these very sad circumstances that this revised version is
being submitted. Had we met as planned, the final version might of
course have been somewhat different but in view of the intense
correspondence that we had in the period of several weeks preceding
his death, I feel reasonably sure that this version is close to what
he wanted it to be. In any case, I have done the best I could. Warm
thanks are due to Philippe Nozières for a critical reading of the
manuscript.

\acknowledgements One of us (G. D.) wishes to thank J.C. Davis for
communicating the paper by Kohsaka et al. \cite{ref1} and for
helpful exchanges; we thank W. Harrison, S. Kivelson, Z.X. Shen and
T. Geballe for interesting discussions. Partial support from the
Heinrich Hertz Minerva Center for High Temperature Superconductivity
and the Israel Science Foundation are acknowledged by G. D.
\bibliography{paper}
\end{document}